\documentclass[12pt]{article}	%%
\usepackage{axodraw}

\def\fig#1{Fig.\,\ref{#1}}

\title{\bf At the root of things} 

\author{\bf Palash B. Pal\\ 
\normalsize Saha Institute of Nuclear Physics\\ 
\normalsize 1/AF Bidhan-Nagar, Calcutta 700064, INDIA}

\date{November 2007}

\begin{document}

\maketitle

\begin{abstract}
  
  Modern theories of fundamental interactions describe strong,
  electromagnetic and weak interactions as quantum field theories with
  certain kinds of embedded internal symmetries called `gauge
  symmetries'. This article introduces quantum field theories and
  gauge symmetries to the uninitiated.

\end{abstract}

%%%%%%%%%%
\section{Things behind the things we see}
%%%%%%%%%%
The reader looking at this page must be establishing some sort of
interaction with the marks of ink that define the letters on the page.
How is this interaction being established?  Well, if it is evening, I
assume that there is an electric lamp glowing in the room.  Light
coming out of that is hitting the page, getting reflected, and
entering the reader's eyes.

In short, the interaction is being established through light.

Suppose we now ask, why is the lamp glowing?  When the lamp was
switched on, how did the lamp, sitting a few meters away from the
switch, get that piece of information?  We know the answer to this
question.  There is a wire connecting the switch and the lamp, which
carried an electric current.  So in this case, the connection was
established through electricity.

What happens when we turn on an electric fan?  There are coils of
wires inside a fan.  When an electric current flows through it, it
generates a magnetic field around the coils.  If we put a metallic
ring within that magnetic field, the field induces a rotation on the
ring.  Once you have a rotating something, it is easy to fit a few
blades on it so that it can send ripples in the air around it.  So
here it is the magnetic field which acts as an agent in establishing
connections.

In the second half of the 19th century, James Clerk Maxwell taught us
that these are not independent phenomena.  Light, electricity,
magnetism: they all are governed by a common set of laws.  So we can
summarize the statements made in the previous paragraphs by saying
that two things can interact with each other through the
electromagnetic field.

In the first half of the 20th century, we cracked the mystery 
of atoms.  An atom has a nucleus in some central position, and
electrons going around it.  How do the electrons know that there is a
nucleus somewhere there?  Because the nucleus contains protons and
neutrons, of which the protons carry positive electric charges.  These
charges create an electromagnetic field around them.  The electrons
hover around through this field.  So in this case also, the
electromagnetic field acts as the matchmaker.

Sometimes if two different substances are brought close together, they
react chemically.  What happens in a chemical reaction?  In short,
molecules break up owing to interactions between the atomic electrons,
and the atoms reorganize themselves into new molecules.  Thus, here
also the interaction is electromagnetic.

While I write, I hold a pen in my hand.  How do I do that?  There is
something going on in the atoms and molecules that constitute the
fingers of my hand which allows them to put a pressure on the atoms
that constitute the pen.  It would be hard for me to describe the
details --- firstly because the processes are complicated and secondly
because I am no expert in physiology.  What I can say for certain is
that some kind of interaction between atoms is responsible for my
holding the pen, and these interactions are electromagnetic.  It is
the same story behind most of the things we do --- speaking, walking,
sitting down, chewing our food --- you name it!

But if my pen slips out of my hand and falls on the floor, that's not
due to electromagnetic interactions.  Here the earth's gravitation is
responsible for the phenomenon.  Just as a charged particle or a
magnet sets up an electromagnetic field around it, a massive particle
sets up a gravitational field around it.  Because of the gravitational
field that the earth creates, the pen in my hand came to know the
presence of the earth near it.  So, as soon it slipped out of my hand,
it went down and hit the floor.

The sun produces a gravitational field around it, and planets feel it
and go orbiting the sun.  The stars and galaxies in the sky are
roaming around in various ways, all because of gravitation.  The moon
is encircling the earth.  How does the moon know about the earth?
Well, because of the earth's gravitational field.

We are talking about arguably the most fundamental question of
physics.  How does any object know about other objects?  How does any
object relate to others?  How do objects influence other objects?  How
do objects behave under such influences?

If nothing like this happened, if everything in the universe spent
their lives without any interaction with anything else, there would
have been nothing to discuss in physics.  And if fact, there would
have been no one to discuss physics, or anything else, either.
Because our body is made out of conglomeration of molecules, and the
organization and function of those molecules depend crucially on the
interaction between them.  Without these interactions, nothing could
have formed --- no sun, no planet, no plant, no insect, no nothing.

Things exist, and events happen, because there are interactions.  And
every phenomenon that we can see with our naked eyes or feel with our
other senses can ultimately be explained with only two kinds of
interactions: electromagnetic and gravitational.

%%%%%%%%%%
\section{Things beyond the things we see}
%%%%%%%%%%
If we try to understand phenomena that take place at scales which are
too small for us to see, we realize that only the aforementioned two
kinds of interactions are insufficient.  Take, for example, the
nucleus of an atom.  There are protons and neutrons in a nucleus.
Protons are charged particles, so two protons repel each other.  The
magnitude of this repulsion is much much larger than that of the
gravitational attractive force that exists between them.  It seems
then that any nucleus, except the hydrogen nucleus which has only one
proton, should break apart because of the repulsive force which
dominates.  Why doesn't that happen?

The reason must be that there is some other kind of force that is
attractive between the members of a nucleus, and which is much larger
than the repulsive electromagnetic force between protons.  Any proton
or neutron creates a field of this force around it.  This is called
the field of the {\em strong force}.  Through its effects, i.e.,
through strong interactions, the net force between protons and
neutrons in a nucleus is attractive.  And this is why we can obtain
nuclei which are long-lived.

Does it mean that any nucleus lives happily forever, being bound by
the strong force?  No, that's not the case either.  We know about the
phenomenon of radioactivity.  More correctly, it is a class of
phenomena in which certain nuclei spontaneously break up, emitting
some particles in the process.  There are different types of
radioactivity.  In one type, a neutron in a nucleus decays to give a
proton, an electron and a particle called the $e$-antineutrino.  How
does that happen?  Strong interactions cannot be responsible for this,
because strong interactions have nothing to do with the electron.
Could it be possible then that the electromagnetic repulsion that we
talked about earlier is somehow dominant in these nuclei?  But that
cannot be the case either, because electromagnetic interactions cannot
change a neutron to a proton, or whatever.

Not strong, not electromagnetic.  Gravitation is negligibly small at
these scales.  So there must be a fourth kind of force.  That force
must be weaker than strong or electromagnetic forces, which is why all
nuclei do not disintegrate.  And for this reason, we can call this
force the {\em weak force}.

It should not be concluded, from what has been said above, that strong
force manifests itself only through neutrons and protons, or weak
force through radioactive nuclei.  There are a host of other phenomena
in which these forces play crucial roles.  Many of them include other
kinds of particles than the ones we have mentioned so far.

In summary, there are four kinds of forces, or four kinds of
interactions.  The strongest one is known, not surprisingly, as
``strong interactions''.  Electromagnetic interactions are next in the
order of strength, followed by weak interactions.  And gravitational
interactions are so negligibly feeble that we will ignore them for
this article.

%%%%%%%%%%
\section{Give and take: the language of quantum field theory}
%%%%%%%%%%
In talking about forces, or interactions, we have used the word
``field'' quite a few times.  That is how forces are felt.  Maxwell
introduced the electromagnetic field in the 19th century.  The
gravitational force was known since Newton, although the equations
governing the gravitational field came to be known only at the
beginning of the 20th century, when Albert Einstein formulated his
General theory of Relativity.  Indeed, about a decade before that, in
1905, Einstein formulated his Special theory of Relativity, and it was
clear from it that no theory of interactions can be complete if only
the force law between two objects is specified: one also needs the
field in order to carry the information about the force from one
object to another which may be sitting a distance away from the
first.

It would therefore follow that we should need field theories for
strong and weak interactions as well.  But before that was found, or
even attempted, a new revolution took place in physics.  It was the
quantum revolution.  In the language of quantum theory, light behaves
as particles dubbed {\em photons}.  The idea was introduced in 1900 by
Max Planck, in an attempt to explain the manner in which hot objects
radiate.  Then, in 1905, Einstein used photons to explain how certain
substances produce an electric current when they are exposed to light,
a phenomenon known as ``photo-electricity''.  The explanation was in
terms of collisions between electrons in the substance and photons in
the light beam.  In the collision, the electron gains energy from the
photon and comes out of the substance.  Since the electrons are
charged particles, an electric current is produced.

If this idea is taken seriously, one should be able to describe other
electromagnetic phenomena in terms of this photons as well.  In
classical field theory, one assumes that energy is transported as
waves.  That description must have a counterpart in which energy would
be transported as particles, or quanta, as they were called in the
early twentieth century.  Besides, the description should be
consistent with the requirements of the special theory of relativity.
Such theories are called {\em quantum field theories}.  The quantum
field theory of electromagnetic interactions, whose formulation began
in the late 1920s and reached its climax in the 1940s, is called {\em
quantum electrodynamics}, or QED for short.

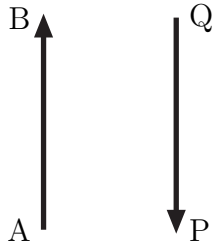
\begin{figure}
\begin{center}
\begin{picture}(60,80)(-30,0)
\SetWidth{2}
\LongArrow(-25,0)(-25,80)
\Text(-30,0)[r]{A}
\Text(-30,80)[r]{B}
\LongArrow(25,80)(25,0)
\Text(30,0)[l]{P}
\Text(30,80)[l]{Q}
\end{picture}
\end{center}
\caption{Two current-carrying wires.  The arrows indicate the
  directions of currents.}\label{sm.f.wires}
\end{figure}
%%%%%%%%%%
So now let us ask: how would QED explain that a current-carrying wire
exerts a force on another current-carrying wire?  Let us look at
\fig{sm.f.wires}.  We have two wires there, carrying currents in
directions shown by arrows.  In such a situation, a repulsive force
will be felt between the wires.  The question is: how?  The two wires
are not touching each other so that information from one wire can pass
on to the other.  How does one wire then know about the other wire?

A classical physicist would follow Maxwell's theory to answer this
question.  He would say that when a current is flowing through the
wire AB, it creates an electromagnetic field around the wire.  Waves
in this field is carrying out the information about this wire.  The
wire PQ falls within this electromagnetic field, and thus it comes to
know about the wire AB.  The wire AB learns about the wire PQ in
exactly the same manner: through the electromagnetic field.

Fine, but we want to understand the same phenomenon from the viewpoint
of quantum theory.  We don't want to invoke the idea of waves in the
electromagnetic field.  We want to talk in terms of photons.  How
should we go about doing it?

Now we will have a different story to tell.  Let us start by thinking
what it means by saying that an electric current is flowing through a
wire.  There are electrons in the substance from which the wire has
been constructed.  Those electrons are flowing.  Now, in course of the
flow, sometimes some electron in the wire AB is emitting a photon.
Maybe this photon travels in the direction where the other wire lies,
and hits that wire.  The photon would then carry the message from the
wire AB to the wire PQ.

I don't mean to say that a specific electron in the wire AB shoots out
only one photon.  There are many electrons flowing through the wire,
and each of them is emitting a photon once in a while.  Not all of
them is emitted in the direction of the wire PQ.  They are being
emitted all around.  So, the space around the wire is teeming with
photons at any given instant of time.  The collection of these photons
is what a classical physicist would call an electromagnetic field.

Among these photons, some travel in the direction of the wire PQ.  The
wire PQ is catching them, or gobbling them up.  It cannot possibly
catch all photons that come its way, but is able to catch some of
them.  Similarly, the wire PQ is emitting a lot of photons, and the
other wire is catching some of them.  The exchange of photons is going
on, establishing the interaction between the two wires.

%%%%%%%%%%
\begin{figure}
\begin{center}
\begin{picture}(120,100)(-60,-50)
\SetWidth{1.2}
\Photon(-30,0)(30,0)27
\ArrowLine(-50,-50)(-30,0)
\ArrowLine(-30,0)(-50,50)
\Text(-55,-50)[r]{A}
\Text(-45,-25)[r]{K}
\Text(-35,0)[r]{B}
\Text(-45,25)[r]{L}
\Text(-55,50)[r]{C}
\ArrowLine(50,-50)(30,0)
\ArrowLine(30,0)(50,50)
\Text(55,-50)[l]{P}
\Text(35,0)[l]{Q}
\Text(55,50)[l]{R}
\end{picture}
\caption{The simplest Feynman diagram depicting interaction between
  two electrons.}\label{sm.f.eescatt}
\end{center}
\end{figure}
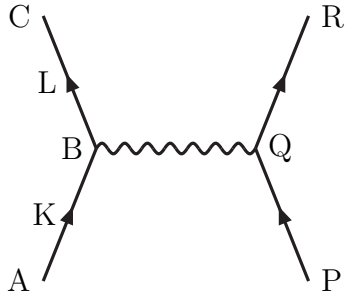
%%%%%%%%%%
It is not really necessary to think about something as complicated as
two wires, with zillions of electrons flowing through them.  We can
think of just two electrons.  Each of the two will somehow feel the
effect of the other.  And the reason would be the same: one electron
would throw some photons which the other one would catch, and vice
versa.

A picture is worth more than a thousand words.  Richard Feynman showed
how to summarize all such words into some simple diagrams.  Look at
\fig{sm.f.eescatt}.  An electron was passing along the path AKBLC.
Along the way, it threw out a photon when it reached the point B.  The
other electron, starting from the point P, caught this photon at the
point Q.

Such pictures are called Feynman diagrams, and they should not be
thought of as a photographic depiction of the real event.  In 
\fig{sm.f.eescatt}, we have not tried to say that the path of the
electron has a sharp bend at the point B.  The photon in the middle is
obviously not traveling along a wavy line.  All lines in the figure
are symbolic: the wavy line represents the photon, the straight lines
represent electrons.  The important message that is given in the
figure is this: only one photon has been exchanged between two
electrons. 

We might ask, why did the electron throw out a photon precisely at the
point B?  Couldn't it have thrown out a photon when it was at K? Or at
L?  What's so special about B?  Similarly, why did the other electron
catch the photon at the point Q?  What's special about this point?

Nothing.  No specialty at all.  In fact, the first electron could
have thrown out the electron at any point.  The probability was there,
all along.  The photon could have been emitted at K or L or any other
point:  it just happened that it was emitted at B.  Similarly, it is
a matter of chance, or probability, that the other electron caught the
photon at the point Q.

%%%%%
\begin{figure}
\begin{center}
\begin{picture}(120,100)(-60,-50)
\SetWidth{1.2}
\Photon(-30,0)(30,0)27
\ArrowLine(-50,-50)(-30,0)
\ArrowLine(-30,0)(-50,50)
\Text(-55,-50)[r]{A}
\Text(-45,-25)[r]{K}
\Text(-35,0)[r]{B}
\Text(-45,25)[r]{L}
\Text(-55,50)[r]{C}
\ArrowLine(50,-50)(30,0)
\ArrowLine(30,0)(50,50)
\Text(55,-50)[l]{P}
\Text(35,0)[l]{Q}
\Text(55,50)[l]{R}
\Photon(-40,25)(40,25)29
\Text(0,-60)[]{\Large (a)}
\end{picture}
\hfill
\begin{picture}(120,100)(-60,-50)
\SetWidth{1.2}
\Photon(-30,0)(30,0)27
\ArrowLine(-50,-50)(-30,0)
\ArrowLine(-30,0)(-50,50)
\Text(-55,-50)[r]{A}
\Text(-45,-25)[r]{K}
\Text(-35,0)[r]{B}
\Text(-45,25)[r]{L}
\Text(-55,50)[r]{C}
\ArrowLine(50,-50)(30,0)
\ArrowLine(30,0)(50,50)
\Text(55,-50)[l]{P}
\Text(35,0)[l]{Q}
\Text(55,50)[l]{R}
\Photon(-40,-25)(40,-25)29
\Text(0,-60)[]{\Large (b)}
\end{picture}
\hfill
\begin{picture}(120,100)(-60,-50)
\SetWidth{1.2}
\Photon(-30,0)(30,0)27
\ArrowLine(-50,-50)(-30,0)
\ArrowLine(-30,0)(-50,50)
\Text(-55,-50)[r]{A}
\Text(-45,-25)[r]{K}
\Text(-35,0)[r]{B}
\Text(-45,25)[r]{L}
\Text(-55,50)[r]{C}
\ArrowLine(50,-50)(30,0)
\ArrowLine(30,0)(50,50)
\Text(55,-50)[l]{P}
\Text(35,0)[l]{Q}
\Text(55,50)[l]{R}
\Photon(-40,25)(40,25)29
\Photon(-40,-25)(40,-25)29
\Text(0,-60)[]{\Large (c)}
\end{picture}
\end{center}
\caption{More complicated Feynman diagrams depicting interaction between
  two electrons.}\label{sm.f.multi}
\end{figure}
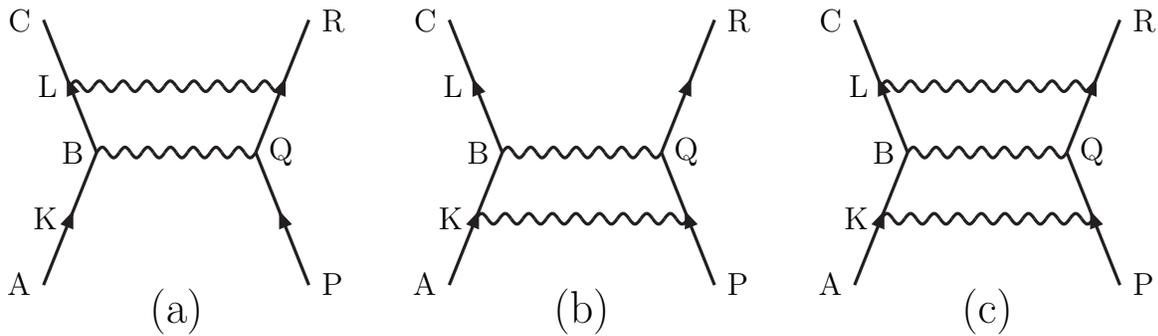
%%%%%
Since the probability exists, could it not have happened that one
photon was emitted at B, and another one at L, and the other electron
caught both of them?  Well, yes, it could have, and Feynman would have
represented it by the diagram of \fig{sm.f.multi}a.  The second photon
could have been exchanged between two different points as well, as
shown in \fig{sm.f.multi}b.  Note the word ``exchanged''.  We are no
longer saying which electron emitted a photon and which one caught it.
In all these pictures, we can also think of the photon being emitted
by the second electron and caught by the first electron.  Many such
exchanges can also take place, as shown in \fig{sm.f.multi}c.  While
calculating the force between two electrons, all such diagrams will
contribute.  We will have to add all of them up in order to obtain the
full interaction between two electrons.

Planck and Einstein, in the first few years of the twentieth century,
showed how the idea of light quanta, or photons, can explain various
phenomena involving light, or radiation in general.  Take, for
example, the case of the photoelectric effect, which happens because
of scattering between light and electrons.  So light, or
electromagnetic field, was the actor in that play.  And now we see
that the electromagnetic field can do more: it can function like the
director of a play, who remains behind the wings, but decides how
different actors should interact with one another.  In other words,
now we can describe all properties of an electromagnetic field in
terms of photons.  This language, or manner, of description is called
{\em quantum field theory}.  In this language, an electromagnetic
field is a collection of many many photons.  Or, turning this around,
we can say that a photon is the quantum, or the particle,
corresponding to the electromagnetic field.

What would have happened if, instead of electrons, we had two
different particles?  We said that at any instant, there is a
probability of the electron emitting a photon.  The same would be true
for any other charged particle.  Suppose we consider two $d$-quarks,
which are some kind of particles whose charge is one-third that of the
electron.  In the language of quantum field theory, it means that at
any instant, the probability of a $d$-quark's emitting a photon would
be one-third the probability of an electron emitting a photon.  The
probability of catching a photon also suffers from the same factor.
Thus, if we think that the solid lines in \fig{sm.f.eescatt}
correspond to $d$-quarks rather than electrons, we would have
one-third of a chance that the quark will emit a photon at the point
B, and a further one-third of a chance that this photon will be
absorbed at the point Q.  So, the process in its entirety would have a
probability of $\frac13 \times \frac13$, or $\frac19$, compared to the
same process with electrons.  The repulsion between two $d$-quarks
would therefore be one-ninth that of the repulsion between two
electrons.  This is Coulomb's law: the magnitude of the force between
two charged particles is proportional to the product of the charges of
the particles.  More complicated diagrams, like those appearing in
\fig{sm.f.multi}, would yield different ratios, but then the
contribution of these diagrams are so small to begin with that they
hardly matter.

We seem to be getting back all results of classical electromagnetic
theory through this new language, sometimes with some small
corrections which went unnoticed in the classical version.  Let us now
ask the question that would prove extremely important in what follows:
how do we obtain the law of conservation of charges in the language of
quantum field theory?

The answer is very simple.  Let us go back to \fig{sm.f.eescatt} one
more time.  What is happening at the point B?  An electron emits a
photon.  The charge of the electron does not change in the process of
course.  Thus, no change of charge will take place in the process if
photon is considered to have zero charge.  No problem with charge
conservation if an electron emits a photon.  No problem if an electron
absorbs a photon.  No problem if the charged particle is not electron
but something else.  No problem if the particle emits a hundred
photons and absorbs seventeen.

%%%%%%%%%%
\section{Symmetry, or the wonderful confusion of being}
%%%%%%%%%%
In the early twentieth century, the renowned mathematician Emmy Noether
showed that conservation laws are intimately connected with
symmetries.  For example, the law of conservation of momentum can be
derived from translational symmetry of space, i.e., the hypothesis
that no point in space is special, and we can set up the origin of a
co-ordinate system anywhere we please, with identical consequences.
Energy conservation can be derived from homogeneity of time, which is
the hypothesis that any instant of time is equivalent to another.

In the same spirit, we can ask, which is the symmetry that is related
to electric charge conservation?

Quantum theory considers particles and waves as complementary
descriptions of any object.  We mentioned that electromagnetic
radiation, which used to be considered as waves in classical physics,
was described as a collection of particles called photons in the
language of quantum theory.  Likewise, quantum theory provided a
description of electrons and other particles of matter in terms of
their associated waves, called {\em matter waves} in general.

By waves, we mean some quantity that is varying in space and time.
When we have waves on the surface of a pond, it is the water level
which varies.  For electromagnetic waves, the electric field, or
equivalently the magnetic field, varies in space and time.  For matter
waves, the similar quantity is almost universally denoted by the greek
letter $\psi$ (psi).  But there is a difference.  The height of water
level or the electric field are represented by ordinary numbers.  The
matter wave amplitude $\psi$, on the other hand, is represented by
somewhat more complicated things called {\em complex numbers}.  For
the purpose of our discussion, we can think of a complex number as an
arrow on a piece of paper, i.e., an arrow in two dimensions.  These
are not physical dimensions like length, breadth or height; these are
some hypothetical directions.  The value of $\psi$ at any point at any
instant can be represented by an arrow at that point at that instant.
The length of the arrow would represent the magnitude of $\psi$, and
the direction of the arrow would represent the direction of $\psi$ in
that hypothetical space embodying complex numbers.  The length is
called the {\em modulus} of the complex number, and the direction its
{\em phase}.

What does $\psi$ mean?  If we consider the $\psi$ of some kind of
particle at a point, the square of the length of the associated arrow
will give probability of finding the particle there.  Good, but not
enough.  We also need to find out the physical implications of the
direction of the arrow.

%%%%%
\begin{figure}
\def\ABline#1#2{\DashLine(5,90)(180,#1)7 \DashLine(5,10)(180,#1)7 
\Text(184,#1)[l]{\tiny\bf #2}}
\begin{center}
\begin{picture}(200,100)(0,0)
%\SetWidth{1.2}
\Text(0,90)[r]{A}
\Text(0,10)[r]{B}
\ABline{64}1
\ABline{57}2
\ABline{50}3
\ABline{43}4
\ABline{36}5
\SetWidth{2}
\Line(181,80)(181,20)
\Text(181,10)[]{C}
\end{picture}
\end{center}
\caption{Two streams of electrons falling on a
screen.}\label{sm.f.2streams}
\end{figure}
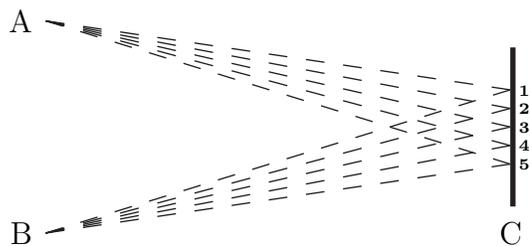
%%%%%
For that, let us suppose that two streams of electrons are coming from
two points A and B, falling on a screen, as seen in
\fig{sm.f.2streams}.  If only the beam from A came to a point, the
matter wave amplitude would have been $\psi_A$.  If it came only from
B, the amplitude would have been $\psi_B$.  When both come together,
the amplitude would then be the sum of the two, i.e.,
$\psi=\psi_A+\psi_B$.

%%%%%

%
\begin{table}[b]
\caption{Arrows represent the phases of electrons coming to different
  points shown in \fig{sm.f.2streams}.  The source points for the
  electrons have been shown as a subscript on $\psi$.  A dot on the
  rightmost column represents a cancellation between the two
  contributions, whereas a double-lined arrow represents
  reinforcement.}\label{sm.t.psi}
\begin{center}
$
\begin{array}{r|ccc}
\hline
& \psi_A & \psi_B & \psi_A + \psi_B \\ 
\hline
1 & \rightarrow & \rightarrow &\Rightarrow\\
2 & \downarrow & \uparrow & \cdot \\
3 & \leftarrow & \leftarrow & \Leftarrow \\
4 & \uparrow & \downarrow & \cdot \\ 
5 & \rightarrow & \rightarrow & \Rightarrow\\
\hline
\end{array}
$
\end{center}
\end{table}
%%%%%
But we have to remember that these objects that we called $\psi$ or
$\psi_A$ or $\psi_B$ are not ordinary numbers.  They are complex
numbers or arrows.  So, if at a point we have the arrows corresponding
to $\psi_A$ and $\psi_B$ which are equal in length but opposite in
direction, the effects of the two would cancel at that point and the
sum, i.e., $\psi$, would be zero at that point.  It would mean that at
that point, there is no way we can find an electron, no matter how
hard we look for it.  If we look at Table~\ref{sm.t.psi}, we find that
such are the cases at the points marked 2 and 4.  On the other hand,
if the arrows corresponding to $\psi_A$ and $\psi_B$ point in the same
direction somewhere, they would reinforce each other, and the
probability of finding an electron would be high there.  This is what
happens at the points 1, 3 and 5 of Table~\ref{sm.t.psi}.  If the
directions of the two arrows are neither the same nor the opposite, the
sum will be somewhere in-between.  In summary, if we have two streams
of electrons falling in a region, there will be points where we will
see a lot of electrons, and other points where we will see less, and
some points where we will see none.  Such a phenomenon is called {\em
interference}, and is expected of any wave.  It was known to happen to
light waves for a long time.  For matter waves, the evidence was
obtained in the first half of the twentieth century.

Let us think about the whole thing.  Suppose we have performed such an
experiment.  We found out where the probability of getting electrons
was zero.  We know that at those places, the arrows for the two
streams were in opposite directions.  In more formal language, we know
that the phases were opposite.  True, but do we know the individual
phases?  The answer is `no'.  We know that in this hypothetical space
where phases are like directions, if $\psi_A$ pointed along the
direction of 10 on the face of a clock, $\psi_B$ pointed towards 4.
If $\psi_A$ pointed towards 6, $\psi_B$ towards 12, and so on.  We
don't know more than that.  Similarly, at places where the probability
of getting electrons was the highest, we know that the arrows
corresponding to $\psi_A$ and $\psi_B$ pointed to the same direction,
i.e., the phases were the same.  But we don't know what that phase
was.

Imagine that we were all asleep at a time when some genies appear and
rotate the directions of these arrows everywhere in the universe.
Would it make any difference?  If we have two arrows and we rotate
both of them by the same amount, the angle between them would not
change.  If they used to be back-to-back, they would remain so.  If
they were in the same direction, they would continue to be in the same
direction after the genies' work.  That would mean that the
probabilities of finding the electron at different points would remain
the same even after this change.  We would not be able to suspect that
the genies have made some change.

If a change of something does not produce any change of something
else, that is called a {\em symmetry}.  We have talked about the
translational symmetry of space and time earlier.  Here we are
encountering a symmetry with the phases.   The relative phase between
two streams of electrons is important, and we can see its
consequences.  But if all the phases are changed by the same amount,
that does not have any effect on the physical universe.

%%%%%%%%%%
\begin{table}

\caption{Same as in Table~\ref{sm.t.psi}, except that an extra half
  rotation has been introduced for electrons coming from the point
  A.}\label{sm.t.psi'} 
\begin{center}
$
\begin{array}{r|ccc}
\hline
& \psi_A & \psi_B & \psi_A + \psi_B \\ 
\hline
1 & \leftarrow & \rightarrow & \cdot \\
2 & \uparrow & \uparrow &  \Uparrow \\
3 & \rightarrow & \leftarrow & \cdot \\
4 & \downarrow & \downarrow & \Downarrow \\ 
5 & \leftarrow & \rightarrow & \cdot \\
\hline
\end{array}
$
\end{center}
\end{table}
%%%%%
But wait, there is a caveat!  The phases must be changed by the same
amount everywhere.  If instead we change the phase by different
amounts at different places, the result will be appreciable.  For
example, let us go back to the two streams of electrons shown in
\fig{sm.f.2streams}.  And suppose we change the phase of the stream
coming from A by a half turn, doing nothing to the stream at B.  In
Table~\ref{sm.t.psi'}, we have shown what will happen now when the two
streams meet.  Previously, the arrows at the points 2 and 4 were
back-to-back.  Now, they would be pointing in the same direction.  In
practical terms, it means that now we would get maximum number of
electrons at a place where we failed to find any electron earlier.

Obviously, the genies need not do something as dramatic as giving a
half turn to the phase at one point in order to be felt.  As long as
they deviate by any small amount from exact equal changes of phases
everywhere, there will be a difference in the interference pattern.
Turning things around, we can say that as soon as we see a change of
interference pattern, we would know that the phase has changed
somewhere: the frivolity of the genies would be exposed.

But these genies do not want to be exposed, so they have arranged a
deep conspiracy.  The point is that, if an electron had emitted or
absorbed a photon at the point A, that also would have changed the
phase of the electron.  That would have caused changes in the
interference pattern if that electron had met another electron
subsequently. 

It means that, if we see a difference in the interference pattern, we
cannot immediately conclude that the arrow of $\psi$ has been rotated
somewhere.  We should be aware that the difference can just as well be
caused by the electron emitting or absorbing a photon on the way.

So the bottom line is the following.  Earlier, we said that if the
genies changed the phases of everything by the same amount everywhere,
we could not have known that.  Now, we find that even if the phases
are not changed by the same amount everywhere, there is no way for us
to know that.

This kind of calculated confusion is called {\em gauge symmetry}.  The
name is bad, and makes no sense, because the word `gauge', in English,
means a measuring instrument, as in `rain gauge'.  But if a
meaningless concoction get universal acceptance, we cannot but go
along with it and think, ``what's in a name!''  Certainly a name such
as {\em phase symmetry} would have been much more appropriate, but who
is listening?

Anyway, let's go back to the question that appeared near the beginning
of this section.  It was a question about the symmetry behind the law
of conservation of charge.  Well, the answer should be obvious now.
The symmetry behind this conservation law is the gauge symmetry that
we described in this section.

%%%%%%%%%%
\section{The same wine in a new bottle}
%%%%%%%%%%
Let us repeat what we just said about gauge symmetry.  The genies want
to change the phases, or the arrows.  If they could do that by the
same amount everywhere at the same instant, we could not have possibly
noticed their work.  But that's easier said than done!  Just imagine:
they will have to change the phase in Calcutta, in Hyderabad, in
Paris, in Abidjan and in Hanoi, all by the same amount, at the same
time.  They will have to do the same behind the clouds, near the sun,
away in the galaxies.  Oh, that's too much even for a genie!  In a
more serious tone, we can say that the tenets of the special theory of
relativity does not even allow such an operation.

Of course there is nothing against changing the phase in a small
region.  The genies can do that.  But they are afraid that we will get
to know what they are doing.  So they have devised a particle called
`photon'.  Because there are photons, we cannot really tell whether
the phases are being rotated.

Of course, as we learned earlier, talking of photons is talking of
electromagnetic interactions.  Thus we can say that the
electromagnetic interactions are results of gauge symmetry.  It is a
fa\c{c}ade to hide the undercurrents of phase rotations.

We are saying the same thing that we said in the last section, but
from a different point of view.  This is the way that two physicists,
Chen-Ning Yang and Robert Mills, described things in 1954.  With this
new way of looking at things, they could generalize the idea to other
kinds of symmetries and hinted that one should try to explain other
interactions with such generalizations.

There are three other kinds of interactions, as we described earlier.
Barely about a decade and a half after the Yang-Mills prescription, it
was seen that weak interactions can be understood through gauge
symmetries.  And then, in 1974, it was realized that strong
interactions can also be explained the same way.  The gauge theory of
strong, weak and electromagnetic interactions constitute what is known
as the {\em standard model} of interactions.

Notice that we have left out gravitation.  We will comment on it
later.  Right now, our aim should be to try to understand the standard
model, i.e., to understand how gauge theories helped understand strong
and weak interactions.  Historically the mystery of weak interactions
was cracked earlier, as we just described.  But we will take an
anachronistic approach and describe strong interactions first, for
reasons to become obvious as we proceed.

%%%%%%%%%%
\section{The pillars of strength}
%%%%%%%%%%
Not everything interacts via strong interactions.  Said another way,
strong interactions cannot affect all kinds of particles.  It can
affect protons and neutrons, both of which are constituents of atomic
nuclei and are therefore collectively called {\em nucleons}.  It can
affect many other kinds of particles, like pions or delta particles.
All these particles are collectively known as {\em hadrons}.

The idea took its root in the early 1960s that these hadrons are not
fundamental particles.  They have another level of substructure, i.e.,
they are made of something more minute.  These minute objects are
called {\em quarks}.  It was conjectured that the proton consists of
two up (or $u$) quarks and two down (or $d$) quarks.  For the neutron,
the tally is opposite: two $d$ quarks and one $u$ quarks.  The scheme
can work if the electric charge of the $u$ and the $d$ quarks is
$\frac23$ and $-\frac13$ that of the proton, respectively.

The idea of quarks brought about great simplification in the task of
understanding hadrons.  For example, the same $u$ and $d$ quarks could
explain the occurrence of many other hadrons, including pions and
delta particles that we mentioned a little while ago.  It was found
that there were four kinds of delta particles: with charges 2,1,0 and
-1 in units of the proton charge, represented usually by the symbols
$\Delta^{++}$, $\Delta^+$, $\Delta^0$ and $\Delta^-$.  With the
charges of $u$ and $d$ quarks mentioned above, it is easily seen that
the combinations $uuu$, $uud$, $udd$ and $ddd$ would fit the bill
exactly for the delta particles.  To understand the structure of all
hadrons that have been discovered so far, we need four more quarks.
That's six quarks in all.

A question that arises is this: how can, say, the combination $udd$
represent both neutron and $\Delta^0$?  Or $uud$, for that matter,
which seems to represent both the proton and a delta particle of the
same charge.  The solution of this apparent mystery lies in the fact
that protons and neutrons have spin-$\frac12$, whereas the spin of the
delta particles is $\frac32$.  Spin, or inherent angular momentum, can
take only integral or half-integral values when measured in a certain
unit, which is what we are using here.  In this preferred unit, each
quark has a spin equal to $\frac12$.  While adding up the spins of
individual quarks, we need to remember that the direction of the spin
is important as well as the magnitude.  So, if we have two of the
quarks pointing in a certain direction but the third in the opposite
direction, the sum total of the three spins would be $\frac12$, which
can be the case with the neutron or the proton.  On the other hand, if
all quarks have their spins pointed in the same direction, the total
spin will be $\frac32$, which is what the delta particles have.

This explains the difference between the nucleons and the delta
particles, but it creates a new problem.  In order to understand
atomic structure, Wolfgang Pauli proposed a hypothesis called the {\em
exclusion principle}.  It asserts that in an atom, two electrons
cannot occupy the same state, i.e., cannot have the same combination
of energy, angular momentum, and a few other things.  The same
principle was applied to nucleons in nuclei, and was successful.  So
there was an expectation that any particle whose spin is $\frac12$
would obey this exclusion principle.

Now, quarks have spin equal to $\frac12$.  And what do we see if we
look at them?  Let us look at the particle $\Delta^-$.  It contains
three $d$-quarks.  According to the exclusion principle, the three
should be in three different states.  But we said a little while ago
that the spin of the deltas in $\frac32$, which can be obtained if all
three quarks have spins pointed in the same direction.  So, as far as
spin is concerned, there is no difference between the quarks in the
$\Delta$ particles.  There is no difference in their orbital motion
either.  How does exclusion principle work then in this case?

It cannot, obviously, unless we assume that we have not mentioned
everything that is required to specify the state of a quark.  If we
use all quantities that are required to specify the state of an
electron in an atom, then, as we saw, exclusion principle goes down
the drain.  For quarks in a hadron, let us assume that there is an
extra quantity which needs to be specified, and let us call this
quantity {\em color}.

Quarks can come in three colors: red, blue and green.  As I say that,
let me warn the reader that I don't mean that some quarks share the
same visual characteristic as the setting sun, some the autumn sky,
and some the leaves on a tree.  That's not what we mean by `color'
here.  This `color' is a new property of matter, and has no connection
with the sense in which we use the word in everyday language.

Whatever property it is, it saves the exclusion principle for us.  As
we said, a $\Delta^-$ contains three $d$-quarks.  The three live in
the same state as long as one does not think of color.  And what
about color?  Well, one of the quarks is red, one is blue, and the
other green.  If we include color as we have to, this ensures that
each quark is in a different state.  Same thing can be said about
$\Delta^{++}$.  It contains three $u$-quarks, but each with a
different color.

Now consider there are genies who are trying to confuse us about this
novel property called color.  They are changing the colors of
everything that is colored.  Are we going to know about it?  If they
change all colors consistently at the same time, we would not know.
In a $\Delta^-$ particle, if the genies changes the red quark to blue,
the blue quark to green, and the green to red, there would still be
one red, one blue and one green quark in the $\Delta^-$, and we would
not face any problem with the exclusion principle or anything else.

But we discussed earlier that these genies are not that efficient.
Rather, they cannot be.  They cannot change the colors of all quarks
everywhere in the same way.  Suppose their activities have been
limited to the place where there used to be a red quark in a
$\Delta^-$, and they have changed it to blue.  Since there was a blue
quark to start with, this change would cause a problem with the
exclusion principle.  And if that happens, we would know what the
genies have been trying to do surreptitiously.

\begin{figure}
\begin{center}
\begin{picture}(120,100)(-60,-50)
\SetWidth{1.2}
\Gluon(-30,0)(30,0)27
\Line(-50,-50)(-30,0)
\Line(-55,-50)(-35,0)
\Line(-60,-50)(-40,0)
\Line(-30,0)(-50,50)
\Line(-35,0)(-55,50)
\Line(-40,0)(-60,50)
\Line(50,-50)(30,0)
\Line(55,-50)(35,0)
\Line(60,-50)(40,0)
\Line(30,0)(50,50)
\Line(35,0)(55,50)
\Line(40,0)(60,50)
\end{picture}
\caption{Gluon exchange mediates strong
  interactions.  The three solid lines on each side are supposed to
  represent three quarks in a particle like the proton or the
  $\Delta$'s.  The exchanged line represents a
  gluon.}\label{sm.f.gluonexc}   
\end{center}
\end{figure}
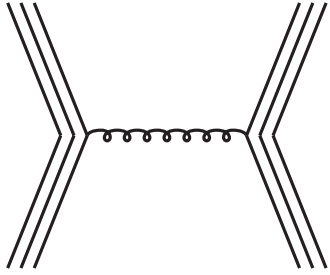

But the genies would not allow us that pleasure.  So they have
invented some new kinds of particles called {\em gluons}, which play
the same role that the photons play in electromagnetic interactions.
A quark emitting or absorbing a gluon can change color.  Thus, a red
quark can change into blue by emitting a gluon, and another quark
might change from blue to red by absorbing the same gluon.  Exchange
of gluons maintain the color, and this is the way that strong
interaction is mediated.  A schematic figure is given in
\fig{sm.f.gluonexc}. 
%%%%%%%%%%
%%%%%%%%%%

This, by the way, is a gauge symmetry, though with a difference.  In
the case of electromagnetism, we commented that the emission or
absorption of a photon does not change the charge of a particle.  In
the present case, we said that the emission of a gluon, for example,
can change the color of a quark.  The emitted gluon carries this
information and dumps it on another quark, which then changes color
accordingly.  Thus, there can be different kinds of gluons.  For
example, one kind can be called $r\bar b$, meaning that if such a
gluon is emitted, it can turn a red quark into a blue quark.  When it
is absorbed, it does the opposite thing of course, i.e., it can turn a
blue quark into red.  Similarly, there would be $b\bar r$ gluons,
$r\bar g$ gluons, and so on.  There will be eight kinds in all.

In the mid-1970s, it was hypothesized that the exchange of these
gluons is the mechanism by which strong interaction operates.  The
gauge theory describing strong interactions in this way came to be
known as {\em quantum chromodynamics}: since `khroma' means color in
greek.

%%%%%%%%%%
\section{Saying things in pictures}
%%%%%%%%%%
What can we say about weak interactions?  Can it be described by a
gauge symmetry as well?  We have explained electromagnetic
interactions by exchange of photons.  Similarly, strong interactions
are mediated by exchange of gluons.  Which particles play the
corresponding role for weak interactions?

Photons and gluons have spin, and its value is 1 in the unit in which
we are specifying all spins.  It was assumed that the mediators of
weak interactions should also have spin 1.  But, unlike photons, these
particles could not be uncharged.  The charge of the hypothesized
particle, in fact, was equal to the charge of the proton.  The
particle did not have a full proper name: only the letter $W$ (for
`weak', presumably) was used to denote it.  Since its charge is
positive like that of the proton, $W^+$ is a more explicit name.  It
was known that, to every particle there must be an antiparticle with
opposite charge.  Thus, corresponding to the $W^+$, there is also a
negatively charged $W^-$.

%%%%%%%%%%
\begin{figure}
\begin{center}
\begin{picture}(120,100)(-60,-50)
\SetWidth{1.2}
\Photon(-30,0)(30,0)27
\Text(0,-5)[t]{$W^-$}
\Text(0,5)[b]{$\longrightarrow$}
\ArrowLine(-50,-50)(-30,0)
\ArrowLine(-30,0)(-50,50)
\Text(-45,-25)[r]{$\mu$}
\Text(-45,25)[r]{$\nu_\mu$}
\ArrowLine(30,0)(70,50)
\ArrowLine(30,0)(50,50)
\Text(35,25)[r]{$e$}
\Text(55,25)[l]{$\bar\nu_e$}
\end{picture}
\caption{Decay of the muon.  The results of the decay are the
  electron, the mu-neutrino ($\nu_\mu$) and the
  $e$-antineutrino.}\label{sm.f.mudecay} 
\end{center}
\end{figure}
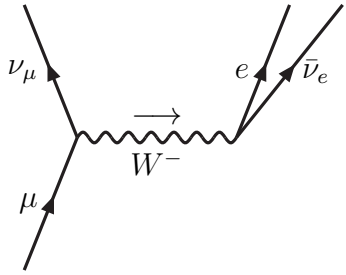
%%%%%%%%%%
Let us see how these particles help us understand the decay of the
muon.  The process has been shown in \fig{sm.f.mudecay}.  The muon has
thrown out a $W^-$ particle.  The charge of the muon, in the unit of
proton charge, was $-1$, same as the charge of the $W^-$.  Therefore,
after emitting the $W^-$, the muon cannot remain a muon: it must turn
into some uncharged particle.  This is the mu-neutrino or $\nu_\mu$.
And the $W^-$, after a while, has turned into an electron and another
uncharged particle, called $e$-antineutrino.  Thus, one obtains three
particles in the decay of a muon.

%%%%%%%%%%
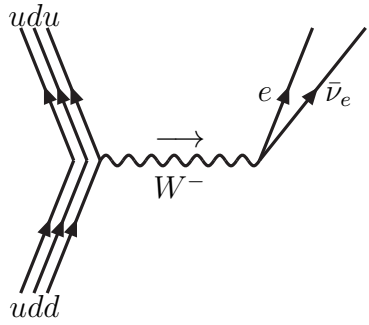
\begin{figure}
\begin{center}
\begin{picture}(120,100)(-60,-50)
\SetWidth{1.2}
\Photon(-30,0)(30,0)27
\Text(0,-5)[t]{$W^-$}
\Text(0,5)[b]{$\longrightarrow$}
\ArrowLine(-50,-50)(-30,0)
\ArrowLine(-55,-50)(-35,0)
\ArrowLine(-60,-50)(-40,0)
\Text(-55,-55)[]{$udd$}
\ArrowLine(-30,0)(-50,50)
\ArrowLine(-35,0)(-55,50)
\ArrowLine(-40,0)(-60,50)
\Text(-55,55)[]{$udu$}
\ArrowLine(30,0)(70,50)
\ArrowLine(30,0)(50,50)
\Text(35,25)[r]{$e$}
\Text(55,25)[l]{$\bar\nu_e$}
\end{picture}
\caption{Decay of the neutron.  The results of the decay are the
  proton, the electron and the 
  $e$-antineutrino.}\label{sm.f.betadecay}
\end{center}
\end{figure}
%%%%%%%%%%
What happens in the case of $\beta$-radioactivity?  Now we need to
look at \fig{sm.f.betadecay}.  Basically, $\beta$-radioactivity means
the decay of a neutron into a proton, an electron and an antineutrino.
The neutron contains three quarks, one $u$-quark and two $d$-quarks.
If one of these $d$-quarks gets metamorphosed into a $u$-quark, we
will obtain a particle with two $u$-quarks and a $d$-quark, which
would be the proton.  And how can this metamorphosis take place?
Well, through the emission of a $W^-$ particle.  This $W^-$, as in the
example of the muon decay, creates an electron and an antineutrino,
and that is how we obtain $\beta$-radioactivity.

These pictures for weak interactions look very much like the
corresponding pictures for electromagnetic or strong interactions.
Instead of the photon or the gluon, we have the $W$, which is the only
important difference.  But then why are weak interactions weak?

The answer that was forwarded was this: the $W$ particles are very
heavy.  This is in sharp contrast with what we had for strong or
electromagnetic interactions.  Gluons are all massless, so is the
photon: their energy is all kinetic.  Because they are massless, it is
easy to emit and absorb them.  For the $W$, since the mass is large,
the same processes are very much inhibited.

Let us be a bit more explicit.  In \fig{sm.f.betadecay}, we see a
proton and a $W$ particle being produced at a point where a neutron is
being annihilated.  Suppose the initial neutron was at rest.  Its
kinetic energy was therefore zero.  It would still have some energy
just because of its mass, which can be obtained by multiplying the
mass by the square of the speed of light in the vacuum.  For the
neutron, this mass-energy is about 940 MeV.  If energy has to be
conserved, the energies of the proton and the $W$ should also be 940
MeV then.  But this is clearly not possible, since, as we know now,
the mass-energy of the $W$ is roughly 81000 MeV.  Even if we forget
about the energy of the proton and possible kinetic energy of the $W$,
we already have a big mismatch.

In classical physics, this would have spelled impossibility of the
event.  Not so in quantum theory.  Note that the $W$ is not produced
as a physical particle in the process: it only appears as an
intermediate state.  Quantum theory allows for a violation of the law
of conservation of energy for intermediate states which are not seen
in experiments.  Only the probability of such occurrences are small
when the mismatch of energy is large.  In the case we have been
talking about, the mismatch is very large, so the process must be very
rare.  It is very difficult to emit or absorb a $W$ particle.  That is
why weak interactions are weak, processes that occur due to weak
interactions are very rare.

An analogy might help.  Suppose the residents of a locality decided
that if anyone makes a surprise visit at someone else's home and finds
no one at home, the visitor must leave a card carrying his or her
name, so that the residents of that home get to know who visited them
while they were away.

In another part of the world and in another civilization, the use of
paper is unknown: they could write only on stone tablets.  They had
the same idea of leaving a `visiting card', but in their case, they
had to carry stone tablets with them whenever they wanted to pay a
visit to anyone else.

It will be trivial to guess which community of people has more
interaction among its members.  Photons and gluons are like paper
cards, and $W$ particles are like stone tablets.  No wonder that weak
interactions are so feeble!

But we discussed that photons are required by gauge symmetry of
electric charge, gluons are required by gauge symmetry involving
color.  Can we not mandate the $W$ by some similar gauge symmetry?

There is a problem though.  If we set up a gauge symmetry in the
manner that Yang and Mills showed us, and then introduce some
particles as guardian angels of that symmetry, these new particles
ought to be massless like the photon or the gluons.  But we just said
that the $W$ particles are very massive.  Hmm, we have a case at
hand!

%%%%%%%%%%
\begin{figure}
\centerline{\epsfbox{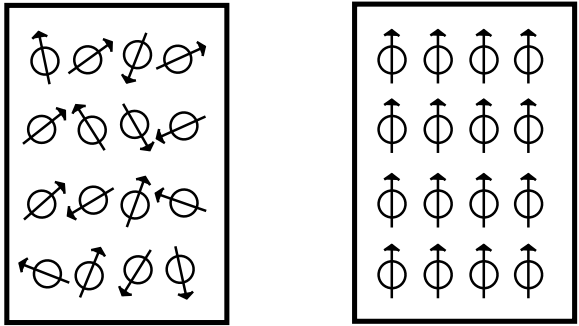}}
\caption{Orientation of little magnets in a piece of
  iron.}\label{sm.f.magnet} 
\end{figure}
%%%%%%%%%%
%%%%%%%%%%
\section{The naughty boys}
%%%%%%%%%%
Indeed, it is true: any gauge symmetry dictates that the gauge bosons
associated with it should be massless.  The question is: do we always
see things that `should' happen?  There are many things which ought to
vanish because of some symmetry, and yet they don't.  Take, for
example, the case of magnets.  If we take a lump of iron and rub it
with a magnet along a specified direction, the lump of iron turns into
a magnet.  Why does that happen?  Let us start from the question why
it does not happen with an ordinary piece of iron.  Each atom inside
the lump of iron is a miniscule magnet.  But normally, the axes of
such magnets are oriented randomly, as shown in the left panel of
\fig{sm.f.magnet}.  If a second piece of iron is brought close to this
piece of iron, each magnet will try to attract this piece in the
direction of its axis.  Since the axes of the magnets are randomly
oriented, so will be the forces, and their effects will cancel.  As a
result, we will see that the piece of iron will not behave as a
magnet.  Once we rub the lump with another magnet, the internal
magnets all get aligned, as shown on the right panel of
\fig{sm.f.magnet}.  In this case, all atomic magnets will pull a
nearby piece of iron in the same direction, and we will conclude that
we have a magnet at hand.

Now imagine a lilliputian scientist sitting inside this magnet.  What
will he or she observe?  All the atomic magnets around the scientist
are pointing in the same direction.  So the scientist might think that
that direction is special compared to others.

We who know the entire story, would not agree.  We think that mother
Nature does not prefer any direction over other.  The atomic magnets
inside the piece of iron are oriented in a particular direction
because we rubbed the piece along that direction.  We have picked one
direction over others by rubbing it in that particular direction.  Had
we rubbed along some other direction, magnetism would have appeared in
that direction.

If we had not chosen any direction and rubbed the piece of iron along
it, the atomic magnets would have remained randomly oriented, and the
total magnetization, summed over all those randomly directed objects,
would have been zero.  Turning things around, we can say that if the
magnetization were zero, the rotational symmetry in the laws of nature
would have been apparent even to the lilliputian scientist sitting
inside the piece of iron.

But symmetries are not always so conspicuous.  Their faces are
sometimes hidden under veils.  This is exactly what happens when the
piece of iron is magnetized.  We could rub the piece along any
direction.  The piece would have developed magnetization along that
direction, no matter what the direction might have been.  There is
symmetry in this respect, and any direction is equivalent.  However,
the fact remains that we rub along some chosen direction, and
magnetization develops along that direction.  As a result, symmetry
has become hidden.  It has made things difficult for the lilliputian
scientist: he or she cannot see that Nature has no preference as far
as directions are concerned.

A similar thing is happening when we, the not-so-lilliputians, are
trying to think about the $W$ particles which mediate weak
interactions.  There is a gauge symmetry which says that the mass of
the $W$ should be zero, just as rotational symmetry says that
magnetization should be zero.  But we are sitting in an universe where
that symmetry is not plainly apparent.  With a hidden symmetry, the
$W$ particle can have mass, just like a piece of iron can have
magnetization.

The physical ideas behind these things developed through the works of
many scientists in the 1950s and the 1960s.  In 1967, Steven Weinberg
created a gauge theory based on such ideas.  A few months later, Abdus
Salam also independently hit upon the same idea.

And this initiated a kind of a revolution in physics.  We said earlier
that the gauge theory of strong interactions was discovered a few
years later.  Thus, with the announcement of the gauge theory of weak
interactions, it was realized that gauge theories are useful in
anything other than electromagnetic interactions.  Of course, the
confidence in this theory did not come overnight.  In fact, most
people doubted the mathematical viability of the theory of hidden
symmetries until, in 1971, Gerhard 't Hooft removed such doubts in a
brilliant set of papers.

There was a strikingly new feature in the theory of Weinberg and
Salam.  They said that the mediators of weak interactions are not only
the charged particles that we have called $W$.  There is another
particle that acts as the mediator, which came to be known as the $Z$.
They are somewhat heavier than the $W$'s.  Like photons, the $Z$
particles do not carry any electric charge.  But, unlike photons, it
can be emitted and absorbed even by uncharged particles like
neutrinos.  Thus, neutrino interactions provide the best testing
ground for this $Z$ particle.  And indeed that was what happened:
some interactions involving neutrinos were observed in 1973 which
could not have been mediated by a charged mediator like the $W$: they
could only be the result of $Z$ mediation.  A decade later, a huge
group of scientists working at CERN (Conseil Europ\'een pour la Recherche
Nucl\'eaire, or the European Council for Nuclear Research) in Geneva under
the leadership of Carlo Rubbia, detected the $W$ and $Z$ particles
directly.  In other words, they observed processes in which the $W$
and the $Z$ participated not as mediators, but as particles present in
the initial or final state of a physical process.

%%%%%%%%%%
\section{Untied knots, unexplored horizons}
%%%%%%%%%%
We have described the basics of the standard model of particle
interactions.  As we saw, the model is based on gauge symmetries.  In
the case of weak interactions, the symmetry is hidden.  For
electromagnetic and strong interactions, the symmetry is apparent.

The model has been remarkably successful in describing particle
phenomena.  Very roughly speaking, we have not seen any particle
phenomena which violates the basic tenets of this model.  There are
numerous situations where the predictions of the model can be
calculated with high precision, and there the results of the
experiments agree with the predictions of the model.

That does not mean that all problems have been solved.  There are
quite a few open ends, and vigorous research is going on to settle
those issues.  Here we list some of them.

First comes the issue of neutrino masses.  Originally when the
standard model was proposed, the neutrinos were assumed to be
massless, because experiments at that time could not establish any
mass of the neutrinos.  Now we know that neutrinos have mass, although
the magnitudes are much smaller compared to the mass of other
elementary particles like the electron.  It is easy to modify the
standard model so that the neutrinos come out to be massive, but it is
generally believed that the modification should hold the clue of the
unusual lightness of neutrinos.  Some interesting ideas have been
proposed in this regard, but it is not clear how to test these ideas
in foreseeable experiments.

If we take the theory of Weinberg and Salam in its original form, even
then we have to admit that some key features of this model have not
been established experimentally.  The theory predicts a spin-0
particle should be left over in the mechanism that provides masses for
the $W$ and the $Z$ particles.  This particle, dubbed the ``Higgs
boson'', has not been observed yet.  

But that's not all that has defied observation.  
The theory of strong interactions bases itself entirely on the idea of
the existence of quarks.  This idea has explained so many experimental
observations that it is hard to disbelieve it.  And yet, it has to be
remembered that it has not passed the acid test for any theory or any
idea: no one has observed a quark in an experiment.

The reason for this might be that the quarks cannot be freed: they are
perennially in bound states which are hadrons.  Such things are not
unheard of.  One pole of a magnet cannot be freed from the other, the
poles always come in pairs.  Perhaps something similar happens for
quarks.  Well, perhaps, but that is a speculation.  No one has shown
that quantum chromodynamics leads us to this conclusion.

To a large extent, the problem lies in the fact that it is very
difficult to calculate the effects of strong interactions when the
quarks are far apart.  Here the word `far' must of course be taken in
context: even the average distance of quarks in a proton would be
considered `far'.  If one wants to free a quark by pulling it apart
from a hadron, one has to pull it to even larger distances, where
calculations are even more difficult and less reliable.  The reason
for such state of affairs in the strength of the interaction.  For
weak and electromagnetic interactions, more complicated diagrams for a
process always give a much smaller contribution compared to the
simplest ones.  For example, consider the interaction between two
electrons, mediated by photons.  We showed some complicated diagrams
in \fig{sm.f.multi}, and a very simple diagram, with only one photon
exchange, in \fig{sm.f.eescatt}.  But, unless one is worried about
very minute corrections, the simplest diagram is all we need.  And,
even if one is worried about some minute corrections, one has to
calculate only a few complicated diagrams, depending on the degree of
minuteness that one is interested in.  For strong interactions, such
rules of thumb do not exist.  Numerical calculations, not dependent
on Feynman diagrams, can be performed on computers, but they have to
make drastic compromises in the nature of the problem in order to
reduce the problem in a calculable form.

We discussed the muon earlier.  It is about 200 times heavier than the
electron, but in all other respects it resembles the electron.  There
is another particle called the tau which is even heavier than the
muon, but has the same properties that the muon and the electron have.
The same structure can be seen among the neutrinos, and among quarks.
The particle physicists say that these are particles from three
generations.  But why are there three generations?  We do not know.

There are aesthetic problems as well.  The standard model, as it is,
contains 19 parameters.  These parameters cannot be calculated: they
have to be determined through experiments.  With them as inputs, we
can find the results of other questions.  But the number 19 does not
make one feel very comfortable.  The number grows once one has to
accommodate neutrino masses.  If you have to give so many inputs to a
theory, it leaves you with a creepy feeling that perhaps you are
missing some deeper understanding which could have cut down on the
number of inputs.

Indeed, one of the persistent dreams of physicists is the idea of
unification.  This is the underlying belief that we will not need
different theories for different interactions: one theory will be able
to describe all of them.  There have been several suggestions
regarding this dream, but no experimental confirmation for any of
them. 

Of course we do not know for sure whether Nature works on a unified
scheme.  But we know for sure that something is obviously missing in
the standard model.  At the very beginning, we said that we will not
consider gravitational interactions because it is negligible at the
scale of elementary particles.  While it is true, it is also true that
we do not know how to describe gravitation in the form of a quantum
theory.  Since the 1980s, string theories have raised the hopes of
describing gravitation.  It is not clear how and whether the other
interactions are contained in such theories.

So there are a lot of things to be done, a lot of ground to cover.  We
have to walk a long way still.  What's more, we do not even know
whether there is an end of the road.  Reflecting on the history of
science, we see that whenever we have cracked a mystery at a certain
level, new mysteries at a new level have been exposed in front of us.
Perhaps the journey is endless, and that is the beauty of the
challenge.\footnote{For the most part, this article is a free
  translation of a chapter from my Bengali book ``ki diye
  somosto-kichu gorha'' (What is everything made of).  I have made
  conscious deviations only in places where the said chapter referred
  to earlier chapters in the book, and in the final section where some
  updating was felt necessary.}

I thank Andrzej K. Wr\'oblewski for pointing out a mistake regarding
the time of publication of Noether's work.

\end{document}